# Rechargeable self-assembled droplet microswimmers driven by surface phase transitions

Diana Cholakova,[1] Maciej Lisicki,[2*] Stoyan K. Smoukov[3*], Slavka Tcholakova,[1] E Emily Lin[3], Jianxin Chen,[3,4] Gabriele De Canio,[5] Eric Lauga[5*], and Nikolai Denkov[1*]

[1]*Department of Chemical and Pharmaceutical Engineering*
*Faculty of Chemistry and Pharmacy, Sofia University,*
*1 J. Bourchier Ave., 1164 Sofia, Bulgaria*

[2]*Institute of Theoretical Physics, Faculty of Physics, University of Warsaw,*
*Warsaw, Pasteura 5, 02-093 Warsaw, Poland*

[3]*Active and Intelligent Materials Lab., School of Engineering and Materials Science,*
*Queen Mary University of London, Mile End Road, London E14NS, UK*

[4]*Department of Applied Chemistry, School of Science, Northwestern Polytechnical University, Xi'an,*
*P. R. China, 710072*

[5]*Department of Applied Mathematics and Theoretical Physics*
*University of Cambridge, Wilberforce Rd, CB3 0WA, UK*

**Corresponding authors for the experimental part:**
Prof. Nikolai Denkov
Corresponding member of the Bulgarian Academy of Sciences
E-mail: nd@lcpe.uni-sofia.bg
Stoyan Smoukov
Email: s.smoukov@qmul.ac.uk

**[#] Corresponding authors for the theoretical part:**
Eric Lauga
E-mail: e.lauga@damtp.cam.ac.uk
Maciej Lisicki
E-mail: maciej.lisicki@fuw.edu.pl

**Abstract**

The design of artificial microswimmers is often inspired by the strategies of natural microorganisms. Many of these creatures exploit the fact that elasticity breaks the time-reversal symmetry of motion at low Reynolds numbers, but this principle has been notably absent from model systems of active, self-propelled microswimmers. Here we introduce a class of microswimmer that spontaneously self-assembles and swims without using external forces, driven instead by surface phase transitions induced by temperature variations. The swimmers are made from alkane droplets dispersed in aqueous surfactant solution, which start to self-propel upon cooling, pushed by rapidly growing thin elastic tails. When heated, the same droplets recharge by retracting their tails, swimming for up to tens of minutes in each cycle. Thermal oscillations of approximately 5°C induce the swimmers to harness heat from the environment and recharge multiple times. We develop a detailed elastohydrodynamic model of these processes and highlight the molecular mechanisms involved. The system offers a convenient platform for examining symmetry breaking in the motion of swimmers exploiting flagellar elasticity. The mild conditions and biocompatible media render these microswimmers potential probes for studying biological propulsion and interactions between artificial and biological swimmers.

**Main text**

Due to their relative simplicity, the natural microswimmers[1,2] and their artificial counterparts[3–5] are convenient systems for studying the complex behaviour of active matter. Indeed, intricate nanomachinery governs movement, from cell shaping and division, to the propulsion of microorganisms in biology.[6] Developments in molecular machines[3] including nanocars[7] foreshadow such life-like complexity in artificial swimmers, but fabrication and integration from the molecular to nano- and micro-scales is non-trivial: "a major challenge is finding robust ways to couple and integrate the energy-consuming building blocks to the mechanical structure".[8] To this end, promising non-biological artificial muscles have already achieved programmable movement[9] and phase-transition driven two-way elastic deformation.[10] Combinatorial methodology to multifunctionality[11] has resulted in self-sensing muscles[12] and the synthetic techniques are applicable to bottom-up synthesis from single molecules into polymer shapes.[13] A number of mechanisms for artificial swimmers have been used, including prominently chemical power[14] – catalytic particles creating bubbles, self-electrophoresis, or releasing slightly dissolving compounds to drive Marangoni flows (e.g. camphor boats).[15] Others are driven by physical effects – thermophoresis[16] or external acoustic,[17] magnetic[18] and electric[19] fields, and serve as a basis for the theoretical understanding of the "active matter".[20]

Three main classes have emerged that are relatively easy to make in large quantities and are therefore accessible for the study of large ensembles of microparticle swarms. The first class, catalytic Janus microswimmers,[5] use catalysts on one side of their surface (e.g. Pt, Pd) to decompose hydrogen peroxide in the surrounding solution, create micro- or nanobubbles, and self-propel while the chemical fuel lasts.

The second major class of artificial microswimmers[21] uses light absorbing Janus particles to induce local heating and asymmetric demixing in a binary lutidine/water mixture, thus generating spatial concentration gradients which induce self-diffusiophoretic motion of the particles. Both types of systems have been extensively studied but the need for toxic peroxide or lutidine prevents the study of such swimmers interacting with biological microswimmers in their native environment.

The third class of microswimmers are compatible with biological media and exploit external light or magnetic fields to create motion. Differential light absorption on relatively easy to fabricate Janus particles has been used for thermophoresis. Such swimmers have achieved ~ 50 μm/s speeds with hollow-particles which allow loading with drugs for targeted delivery.[22] More recently, emulsion-based Pickering immobilization was employed to scale-up and simplify the fabrication of such metal-containing Janus particles, capable of thermophoretic swimming.[23] Their potential beyond drug delivery also extends to therapies such as thrombosis ablation treatment.[24] Magnetic swimmers require more involved microfabrication of intricate screw-type structures[25] which, however, can be controlled in deep tissues without restrictions for transparency or complications of light absorption, thus featuring excellent properties for studies of mixed artificial and biological swimmers. Magnetic micro-robots picked, encapsulated, and delivered cells to locations while protecting them from shear forces,[26] while others were used to capture non-motile sperm cells, propel them and fertilize an egg.[27] Since external magnetic fields synchronize all the swimmer movements and orientation, this class possesses some inherent limitations for studying complexity in active matter.

Still, a minimal non-living model swimmer, which is easy to generate (e.g. by self-assembly) and operates in biologically-compatible media, remains an outstanding experimental challenge. That challenge is so much more difficult, if one wished to assemble a swimmer that could use elasticity for hydrodynamic propulsion, was able to store energy internally and could be recharged.

Here we present a new class of active microswimmers grown via bottom-up molecular self-assembly using only three simple components – alkane oil drops, and water containing a dissolved surfactant. The operating temperature window is tuneable by the choice of oil and surfactant, and small (~5 °C) thermal oscillations in the environment are sufficient to drive and recharge the swimmers, requiring no additional fuel. The experiment requires only an optical microscope with a thermally-controlled sample holder. The compatibility of our system with bacteria and higher organisms[28,29] provides an opportunity to study interactions between artificial and biological swimmers, also in populated swarms.

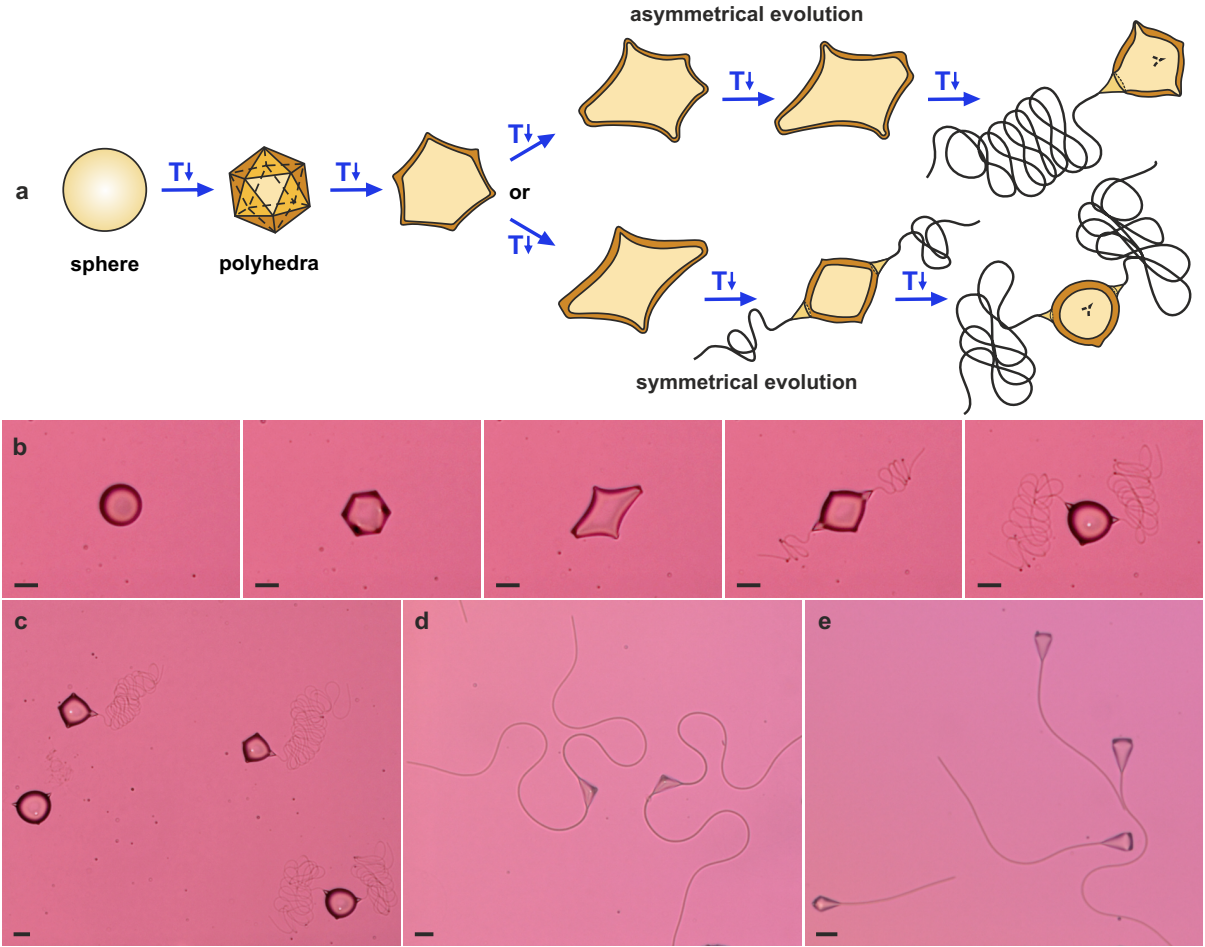

**Figure 1: Emulsion droplets deform upon cooling and eventually form dynamic swimmers with one or two nozzles extruding fibres**. (**a**) Schematics of the transformation of the initial oil drop into swimmer with one or two tails, passing rapidly via series of drop-shape shifts; (**b**) Images of swimmer formation observed experimentally upon cooling of tetradecane drop. (**c**) Microscopy picture of tetradecane swimmers extruding one or two fibres of diameter ≈ 0.5 μm; extrusion rate $U_F$ ≈ 6.5 μm/s for drop extruding two fibres and $U_F$ ≈ 12 μm/s for drops extruding single fibre. The swimmer speed is $U_S$ ≈ 0.25 μm/s for drops extruding two fibres and $U_S$ ≈ 0.5 μm/s for drops extruding single fibre. (**d,e**) Images of pentadecane swimmers extruding fibres of diameter ≈ 2 μm. (**d**) Swimmers, extruding two fibres at rate $U_F$ ≈ 2.3 ± 0.3 μm/s and swimmer speed $U_S$ ≈ 0.45 ± 0.06 μm/s. (**e**) Swimmers extruding one fibre with $U_F$ ≈ 0.85 ± 0.1 μm/s and swimmer speed $U_S$ ≈ 0.28 ± 0.05 μm/s. In all experiments, the alkane drops are dispersed in 1.5 wt. % Brij 58 surfactant solution. Scale bars, 20 μm. The quoted values of $U_S$ and $U_F$ are for the specific drops shown in these images. The relation between $U_S$ and $U_F$ is expressed by eq. (3). The statistically averaged values of the parameter, $c$, entering eq. (3), are presented and discussed in the main text.

Upon cooling, the alkane droplets spontaneously eject thin elastic filaments, which due to viscous friction with the surrounding fluid, push the droplets and induce swimming. Upon subsequent heating of the environment, the filaments retract completely, thus returning the droplets to their initial state and recharging the system. The internal liquid-to-plastic phase transition that occurs on the surface of the oil drops and drives these phenomena is reversible, while the inherent elasticity of the filaments breaks the time-reversal symmetry of the droplet motion (although the latter occurs at low Reynolds numbers) thus generating partially irreversible swimming motion. Developing a detailed elasto-hydrodynamic model of the filament dynamics, we provide quantitative insight into the swimming behaviour of the droplets and highlight some similarities with the beating patterns of flagellated swimmers.[30]

As an illustration we present results obtained with oily drops of alkane (pentadecane or tetradecane), placed in ~1.5 wt. % aqueous surfactant solution (Brij 58) which are cooled at a rate of 0.1 to 1 °C/min, down to ≈ 8 °C for pentadecane ($C_{15}$) or to ≈ 2 °C for tetradecane ($C_{14}$). The cooling results in initial changes in the drop shape which reach quickly a steady-state spheroidal shape with four (or five) "spikes" arranged in the positions of the corners of a tetrahedron (or of a pentagonal pyramid with tetragonal base) at the particle surface, see **Figure 1**. One or two of the spikes transform into nozzles which extrude quickly material from the sphere into long, uniform in diameter filaments (**Figure 1**, **Videos 1-3**). We observed that swimmers with a single tail are formed preferably at lower cooling rates (e.g. 0.1°C/min) whereas the main fraction of swimmers had two tails at a higher cooling rate (*ca.* 0.5°C/min), while drops with one and two tails were observed to coexist in the same sample in the transition range of cooling rates, ca. 0.2 to 0.3 °C.min$^{-1}$. No significant dependence of the number of extruded tails on the drop size was observed, thus excluding a strong influence of the local interfacial curvature for the spikes transformation into fibre-extruding nozzles.

$C_{15}$ and $C_{14}$ droplets extruded fibres with notably different diameters, $d \approx 2.0 \pm 0.2$ μm for $C_{15}$ and $d \approx 0.5 \pm 0.1$ μm for $C_{14}$, independently of whether 1 or 2 fibres were extruded from a given drop (cf. **Figure 1b,c** and **Figure 1d,e** as illustrative examples). For convenience, hereafter we term these fibres "thick" for $C_{15}$ drops and "thin" for $C_{14}$ drops.

The time for which the swimming can be observed in a given system depends mostly on the cooling rate. Depending on the specific oil-surfactant combination, the swimmers are observed in a specific temperature range, e.g. 3-4°C wide for the $C_{15}$ drops. Performing experiments with different cooling rates, we change the duration of the period in which the emulsion temperature falls within this range. For example, at a cooling rate of 0.5°C/min, about 5-10 min is available for swimming, whereas this time is *ca.* 20 min at the lower cooling rate of 0.15°C/min.

In our previous studies, quasi-static shapes were obtained upon slow cooling of emulsion droplets.[31] The experiments showed that these shape transformations (artificial morphogenesis) were driven by the formation of a 2D plastic rotator phase at the drop surface, with somewhat different stability in microconfinement than bulk rotator phases,[32] and with thickness between several and dozens of nanometres, depending on the system.[31,33] This plastic rotator phase is formed at the edges of the deformed drops (i.e. at the edges of the polyhedral shapes, at the periphery of the flattened platelets, and on the surface of the cylindrical protrusions), thus forming a frame of plastic rods which gradually extends with time by incorporating new alkane molecules from the liquid interior of the drops. The molecules in the rotator phase have some rotational freedom and occupy larger volume when compared to the truly crystalline phase of the alkanes – this freedom leads to higher molecular mobility and inherent plasticity of the rotator phases. The energy gain upon the formation of this plastic rotator phase on the surface of the deforming drops is greater than the energy penalty from the expanding interfacial area, where the interfacial energy was measured to be in the range of 3 to 10 mJ/m$^2$.[34] Molecular dynamics tools to model rotator phase transitions were only recently established,[35] but the main stages of the observed drop-shape evolution sequence were explained with a theoretical model which analysed the energy dependence on the droplet shape, with included both the drop surface energy and the energy of the forming plastic frame.[36,37] The observed shapes were interpreted and discussed[38] also as a new type of "tensegrity" (tensorial integrity) structures[39] which acquire mechanical stability by balancing the compression stress of the interfacial tension with the rigidity of the plastic frame forming at the drop surface.

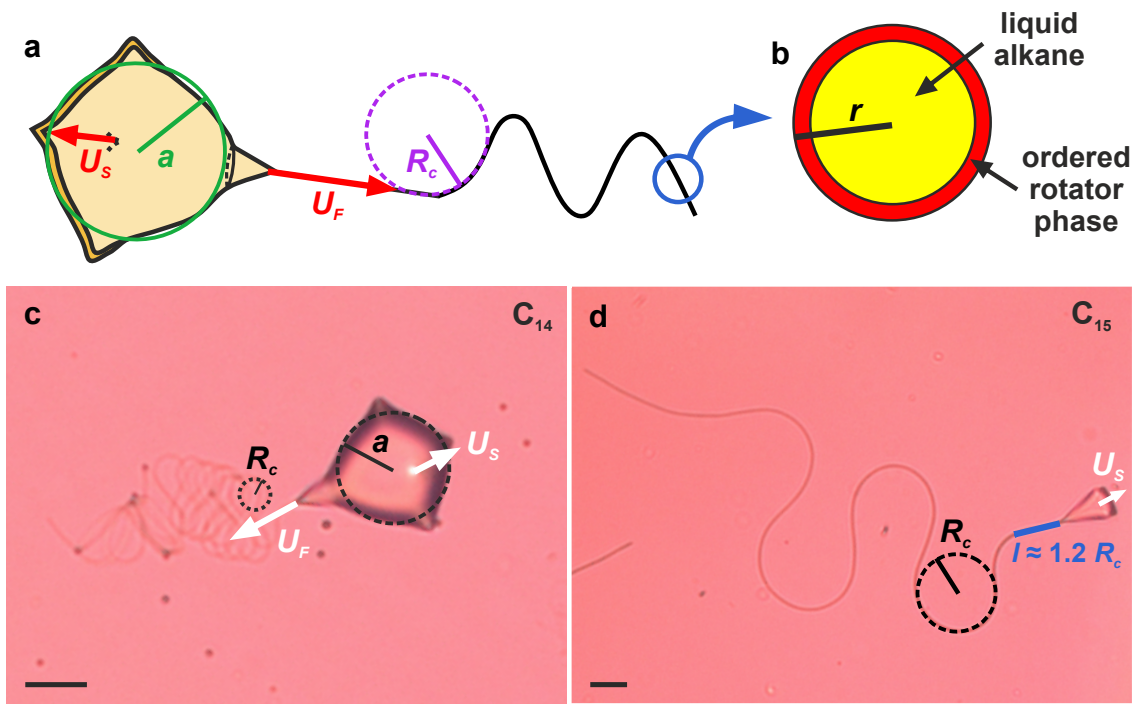

**Figure 2.** (**a**) Detailed diagram of the spiked ball swimmer, noting the effective swimmer radius, $a$, swimmer velocity, $U_S$, fibre extrusion velocity, $U_F$, and the radius of curvature of the first filament bend, $R_c$. The filament radius is $r$ and the plastic shell thickness $\delta$. (**b**) Schematic cross-section of the fibre. In the centre, the fibres are filled with liquid oil, whereas ordered layers of plastic rotator phase are formed on their surface. This rotator phase ensures the fibres' elasticity. (**c-d**) Microscopy images of one-tailed swimmers overlaid with several different quantities measured from the experiment. Swimmer shown on (c) is made from tetradecane oil and the one on (d) from pentadecane oil. Scale bars, 20 μm.

In the new class of oil-surfactant systems described here, the previously observed transformations occur very quickly and the drops rapidly transmute into swimmers (**Figure 1**, **Videos 1-3**). As all the other shape changes are driven by a few layers of plastic phase formed on the surface of the self-shaping drops, we postulate that the elastic filaments have a similar structure. Therefore, the swimmers' actuation is caused by self-assembled thin elastic flagella-like tails, **Figure 2a,** with a shell of thickness $\delta$ composed of alkane molecules ordered in plastic rotator phase, and an interior of liquid alkane, **Figure 2b**. Our experiments showed that two main conditions should be satisfied simultaneously to observe swimmers of this type: (1) the surfactant should be with longer tail than the alkane molecules, so that the surfactant adsorption layers freeze before the drop interior, and (2) the cooling rate should be very low, *ca.* ≤ 1°C, thus allowing the drops to pass through all preceding stages to form a swimmer before their complete freezing.

We quantify the relationships between all key parameters describing the observed swimmers behaviour. The swimming, $U_S$, and extrusion, $U_F$, speeds were measured to increase approximately linearly with the cooling rate which is consistent with faster formation of the plastic rotator phase on the drop surface, **Figure 3a**. Both $U_S$ and $U_F$ are significantly higher for single-tail swimmers compared with the two-tail swimmers at a given cooling rate. This result is also expected, because the rate of plastic phase formation on the drop surface is expected to be the same for a given system and cooling rate, independently of the number of extruded fibres, while this same phase is distributed into one or two fibres in the two types of swimmers. Under equivalent conditions, droplets with smaller initial diameter extrude fibres at a lower rate as compared to bigger droplets, again because bigger drops have a higher surface area from which the rotator phase for the fibres originates. All these experimental trends are captured quantitatively in the model explained below (eq. 3).

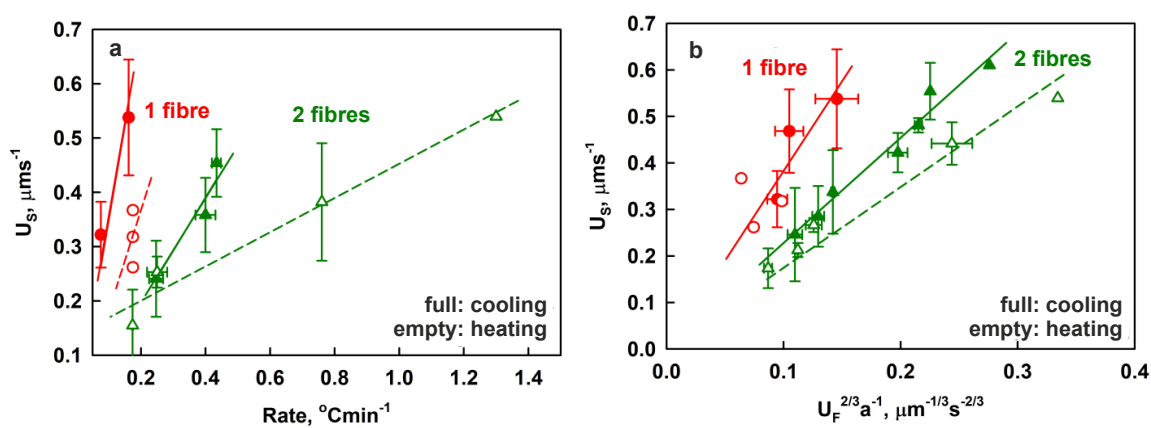

**Figure 3. Droplet swimming speed.** (a) Dependence of the swimming speed, $U_S$, on the cooling rate (full symbols) and heating rate (empty symbols) for $C_{15}$ swimmers in 1.5 wt. % Brij 58 solution, extruding one fibre (red circles) or two fibres (green triangles). (b) Relation between the swimming speed, $U_S$, the fibre extrusion speed, $U_F$, and drop radius, *a*, for the same experimental data, eq. (3). Note that eq. (3) used to construct the plot in (b) is derived for the extruding fibres only – the data with retraction are shown for comparison only.

To clarify the origin of this new type of microswimmer propulsion, we balance the hydrodynamic Stokes drag on the rapidly extruding cylindrical filaments with the drag created by the friction of the propelled mother drop (approximated as a sphere) with the surrounding fluid. The resulting elasto-hydrodynamic model of the swimmers advances the understanding of elastic fibre extrusion dynamics[40–42] and is described in detail in section "II. Theoretical model" in the **Supplementary Information** (SI). Briefly, because of the filaments friction with the viscous medium, we estimate the fibre elastic stiffness from the measured periodicity of fibre buckling with characteristic wavelength, $l$, **Figure 2**. While past calculations only focused on the hydrodynamic friction and the bending of fibres with fixed basis,[40] our theoretical approach builds on a different numerical scheme[41] and fully characterises the free particle swimming using the extrusion of elastic filament, **Figure 2**. The relationship we derive between the swimming speed of a droplet extruding one or two filaments, $U_S$, and the extrusion speed of that filament, $U_F$, is

$$U_S = c U_F \, l/a, \qquad (1)$$

where $a$ is the radius of the main body of the swimmer and $c$ is a dimensionless constant. We treat the filaments as uniform material with characteristic (elasto-hydrodynamic) buckling length, $l$, which depends on the bending stiffness, $A$, the extrusion speed, and the resistance to flow of the filament in the direction parallel to its axis per unit length $\xi_\parallel$ as

$$l = \left(A / \xi_\parallel U_F\right)^{1/3} \qquad (2)$$

The results of our simulations show that the buckling length can be extracted from the radius of curvature $R_C$ of the first buckle, which is measured with high precision from the experimental observations, as $l \approx 1.2 R_C$, **Figure 2**. Combining eqs. (1)-(2) with the theoretical estimates[42] for $\xi_\parallel$ we are able to describe the dynamics of the swimmer using $a$, $l$, $U_F$ and $U_S$ as experimentally accessible quantities, while $c$ and $A$ were determined from eqs. (1) and (2), respectively. We analysed more than 50 swimmers in total, finding the swimming speed $U_S$ to be rather significant, around 1 μm/s, and appears as 10-50 % of $U_F$ of the thick fibres and 3-5 % of the extrusion speed $U_F$ of the thin fibres. From these experiments we determined the bending stiffness $A \approx 210 \pm 60$ Nm$^2$ with $l \approx 25 \pm 5$ μm for the thick fibres, and $A \approx 25 \pm 8$ Nm$^2$ with $l \approx 7 \pm 2$ μm for the thin fibres. Both the one-tailed and two-tailed swimmers are described by this model with the same value of the only material parameter, $A$, a clear indication for the self-consistency of the theoretical approach. This interpretation showed also that the constant $c$ is fairly independent of the materials used and the drop size. Statistically averaging the values of $c$ determined for the individual droplets, we obtained $c \approx 0.142 \pm 0.035$ for swimmers with one filament, which is by $\approx 50$ % higher than $c \approx 0.093 \pm 0.031$ determined for the swimmers extruding two filaments in the same system, probably because the propulsion forces exerted by the two fibres in the two-tail swimmers do not act in the same direction.

The combination of eqs. (1) and (2) yields the following relationship which can be verified experimentally:

$$U_S = \frac{c}{a}\left(\frac{A}{\xi_P}\right)^{1/3} U_F^{2/3} . \qquad (3)$$

Indeed, the plot of $U_S$ vs. $(U_F^{2/3}/a)$ for swimmers of the same composition which presumably have the same values of $A$ and $\xi_\parallel$, gives a straight line, **Figure 3b**. Note that the scaling prediction eq. (3) reflects the intricate coupling between the hydrodynamic propulsion force created by the extruding fibre and its elastic properties which lead to fibre buckling and thus modulate the propulsion force (see Section IV.G in the **SI** for further detail).

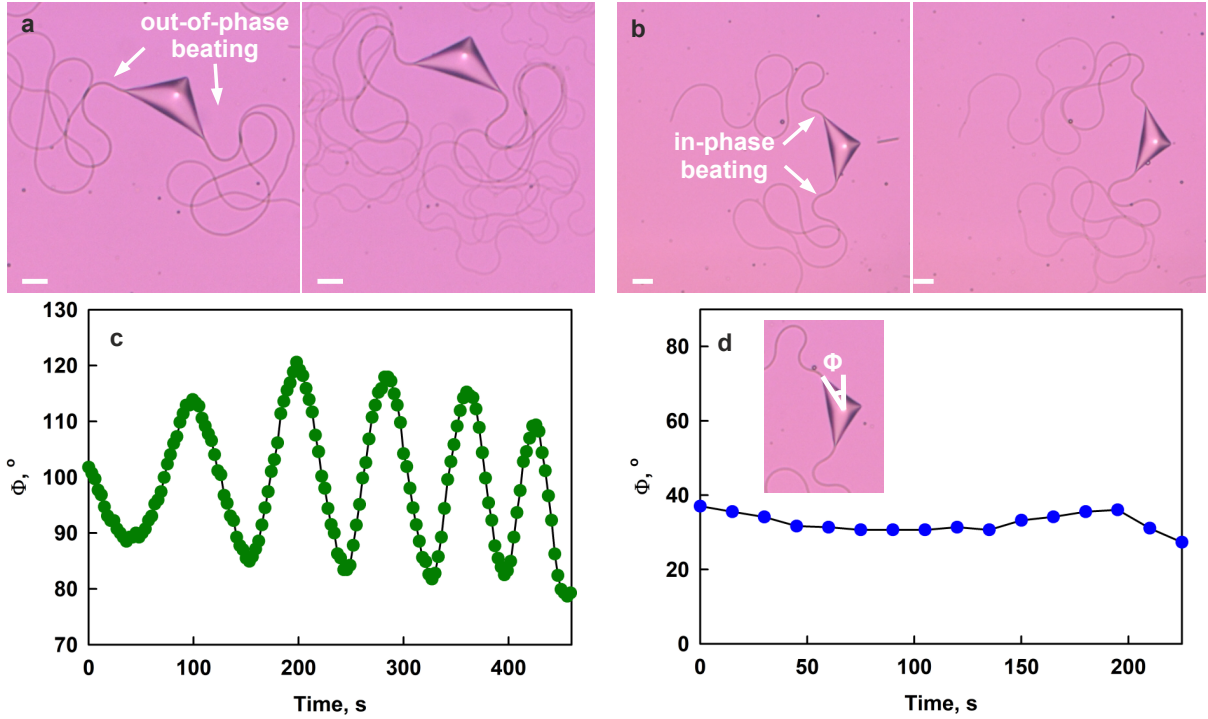

**Figure 4. Kinematics of swimming.** (a-b) Microscopy pictures of pentadecane swimmers dispersed in 1.5 wt. % Brij 58 solution, extruding two fibres. Note that the thickness of the extruded fibres decreases with time due to the molecular rearrangement of the alkane molecules in the fibres, respectively, the length of the fibres increases significantly. (a) The fibres are extruded out-of-phase, see **Video 3**. (b) The fibres are extruded in-phase, see **Video 2**. (c-d) Dependence of the angle Φ with time. This angle is defined as the angle between the extruding tip, the white dot in the centre of the extruding drop and the vertical axis, see inset in (d). For the out-of-phase extrusion it oscillates with time (c), whereas for in-phase extrusion the angle remains almost constant (d). Scale bars, 20 μm. The error bars represent the standard deviations calculated from our data points.

With our model, we also describe quantitatively the undulations of the orientation angle of the swimmers with respect to their direction of motion, see **Figure 4**. Using the experimentally available values of $a$, $l$ and $U_F$ we predict the period and the amplitude of the angle oscillations which agrees very well with the measured data for the last two quantities - see eqs. (S52) and (S58), **Table S2** and the related explanations in Sections IV.F and IV.I in **SI**.

The retraction of the fibres, upon subsequent emulsion heating, is also of high interest, because it defines the reversibility of the process. The retraction is driven by the positive interfacial tension at the oil-water interface, which draws the fibre in to minimise the liquid interfacial area. Unlike the fibre extrusion where the buckling instability coupled with elasticity induced undulation, during retraction both the fibre and droplet are pulled towards the single point of the nozzle resulting in reduced undulating movement of the droplet-filament system. The quantitative comparison of the data for fibre extrusion and contraction shows that the swimmer speeds during extrusion are higher than those upon retraction at the same magnitudes of the cooling and heating rates, see Figure 3a. The slope of the relation between $U_S$ and $U_F^{2/3}/a$ is also somewhat lower for the retraction in the case of two-tail swimmers, see Figure 3b. This difference in the swimming speeds upon cooling and heating, enabled by the fibre elasticity, shows that the swimming is partially reversible. Similar difference is observed also in the simulations, **Video 4** and **Video 5.** Further experiments are needed to quantify more precisely the irreversibility of the observed processes.

The analysis of two-tailed droplets suggests an analogy with biological swimmers. The undulating motion of our swimmers resembles to some extent that of several eukaryotic microorganisms which swim by waving flexible flagella. In particular, the swimmers which extrude two fibres show similarities to the motion exhibited by the biflagellate algae, akin to *Chlamydomonas reinhardtii*, see **Videos 2 and 3**. One question for these biological swimmers concerns the mechanism by which their two flagella synchronise. It had long been thought that the synchrony of beating flagella is achieved through hydrodynamic interaction,[30,43] as seen in artificially driven colloidal systems.[44] The flow fields around two single-flagellated algae have been shown also to lead to concerted motion provided the distance of flow-mediated interactions was sufficiently small.[45] This view has been questioned with increasing evidence that intra-cellular coupling must play a mediatory role in the coordination of beating.[46,47] To test these possibilities with our model system, we studied droplets that were extruding two fibres. By analysing the video records, we observed that the extrusion speeds of the two fibres extruded by a given drop were equal. However, for the various droplets these pairs of fibres were in different relative phases which resulted in different ranges of undulation angles Φ for the main droplets, see **Figure 4** and **Videos 2** and **3**. Unlike the beating patterns observed in *Chlamydomonas* that exhibit changes in the relative phase between the two flagella, our analysis of 30 droplets swimming by extruding two filaments in 12 videos show that, for a given droplet, the phase difference between the extruded filaments remains constant without any sign of synchronization. These results suggest that, at the distances comparable to the sizes of our droplets, the hydrodynamic interactions and the coupling through the body of the droplet are too weak to induce a significant change in the phase difference and fibre synchronisation. One can expect that the internal coupling inside the organism is most likely to play a key role for similarly-sized biological swimmers.

Another analogy with living systems is the emergent ability to harvest energy from changes in the environment (e.g. in the day/night cycles). When we warm rotator phase fibres, often several millimetres long and only a micrometre wide, they retract fully all the way back to the mother drop, see **Videos 6** and **7**. Supplying gentle temperature oscillations of less than 5 °C, which do not lead to oil drop freezing, provides enough energy to recharge the swimmers completely in every cooling/heating cycle and let them swim for multiple cycles, see **Video 8** which shows a droplet in three consecutive extrusion/retraction cycles. The enthalpy of freezing of hexadecane is $\Delta H° \approx 235$ kJ/kg[48] and since $\approx 75\% \approx 175$ kJ/kg of this enthalpy is due to the liquid-rotator phase transition,[48] the potential stored energy in the particles that could be used for swimming exceeds the maximum energy density of a lead-acid battery of ~ 140 kJ/kg (40 Wh/kg).[49]

In conclusion, we present a new class of active, elastic microswimmers produced by simply cooling a 3-component system – oil droplets in aqueous surfactant solution. The swimmers in this class are not restricted to the specific examples presented and discussed in the current paper. We typically observed such active swimmers when surfactants of different types (ionic, nonionic) have saturated hydrophobic tails which are by 1 to 3 C-atoms longer than the alkane molecules and the emulsions are cooled slowly at *ca.* 0.1÷0.5 °C/min. The temperature interval of the swimming behaviour can be tuned by selecting alkanes with appropriate melting temperature in the drops. Our theoretical model has identified the key parameters governing the motion and may inspire new discoveries in active matter. By coupling the buckling instability with filament extrusion we reveal quantitatively the origin of the partial time-irreversibility of this mode of swimming at low Reynolds numbers[28] and provide some insights into the motion of living microswimmers. Note that the nonionic surfactants used in our study are biocompatible and have been applied in various bio-systems,[50–54] although the biocompatibility of our systems with real microorganisms has to be investigated in following studies. We highlight the potential for hydrodynamic studies in the area of active matter by referring to mixed systems of artificial and biological microswimmers which can be explored in diluted or in dense populations to reveal collective effects.

**Acknowledgments:**
This study was funded by the European Research Council (ERC) EMATTER (# 280078) and the Engineering and Physical Sciences Research Council Fellowship No. EP/R028915/1 to S.S. This project has received funding from the European Research Council (ERC) under the European Union's Horizon 2020 research and innovation programme (grant agreement 682754 to E.Lauga). The study received financial support from project № KP-06-DV-4/2019 with the Bulgarian Ministry of Education and Science, under the National Research Program "VIHREN" to N.D. The work has been supported by the National Science Center of Poland grant SONATA to M.L. no. 2018/31/D/ST3/02408. The study falls under the umbrella of European network COST CA17120 Chemobrionics. The authors are grateful to Mrs. Mariana Paraskova (Sofia University) for her help with part of the image analysis and for the preparation of some figures, as well as to the Reviewers of the manuscript who helped to improve considerably this paper with their valuable comments and suggestions.

**Authors contributions:**
D.C. discovered the phenomenon and clarified the experimental conditions under which this new type of swimmers are obtained and can be controlled; D.C., S.T. and N.D. suggested studying the process in more details; D.C. and S.T. designed the experimental part of the study; S.S. designed the part of the study about filament retraction; D.C. performed most of the experiments with respect to fibre extrusion, summarized the obtained results and analysed them (with input from S.T., N.D. and S.S.), while E.Lin, D.C. and J.C. performed most of experiments for fibre retraction (with input from S.S.); E.Lin clarified the experimental conditions for controlled retraction of the tails; S.S. made the first analytical model for the swimming by estimates of sphere and cylinder drag forces; M.L. and E.Lauga developed the theoretical description for the extrusion of the fibre and the motion of droplets; S.S., D.C. and M.L. analysed movies and developed insights to connect dynamic features to material properties of the fibres; M.L. and G.D.C. developed the computer code used in the numerical simulations; S.S. and N.D. prepared the initial manuscript draft; D.C. edited the manuscript and prepared the figures and movies; M.L. prepared the theoretical part of SI; M.L. and E.Lauga edited the manuscripts. All authors critically read the manuscript and approved it.

**Competing interest statement**

None of the authors have competing financial or non-financial interests as defined by Nature Research.

**Data availability statement**

The data that support the findings of this study are available from the corresponding authors upon reasonable request.

The code used in this study is available from the corresponding authors upon reasonable request.

**Methods**

*Preparation of samples and observations*

For preparation of the microswimmers we used emulsions prepared with tetradecane or pentadecane oil dispersed in 1.5 w% of Brij 58 aqueous surfactant solution prepared with deionized water (purified by an Elix 3 module, Millipore, Germany). All chemical substances were obtained from Sigma-Aldrich. The surfactant was used as received and the alkanes (purity 99 %) were purified from surface-active contaminations by passing through a glass column filled with Florisil adsorbent.

Emulsions were prepared using laboratory Microkit membrane emulsification module from Shirasu Porous Glass Technology (SPG, Miyazaki, Japan), working with tubular glass membranes with an outer diameter of 10 mm and a working area of approximately 3 cm$^2$. Membranes with mean pore size of 5 μm and 10 μm were used.

For the optical observations, sample of the prepared emulsion was placed in a glass capillary with rectangular cross sections (dimensions: 1 or 2 mm width; 100 μm height; 50 mm length) and the capillary was placed into a custom-made cooling chamber connected to a cryo-thermostat (Julabo CF30) allowing one to control precisely the temperature. To ensure correct measurement of the temperature, a calibrated thermocouple probe was inserted in a next orifice and the temperatures were recorded during the experiments. All observations were made in transmitted cross-polarized white light. Long-focus objectives x10, x20 and x50 were used to observe the drops upon sample cooling. An additional λ plate (compensator plate) was situated between the polarizer and the analyser, the latter two being oriented at 90° with respect to each other. The λ plate was oriented at 45° with respect to both the analyser and the polarizer. Under these conditions, the liquid background and the fluid objects have magenta colour. Observations were performed with AxioImager.M2m microscope (Zeiss, Germany). The cooling rates applied were varied between 0.05°C/min and 1°C/min and heating rate – between 0.1°C/min and 3°C/min.

*Procedure for Video analysis*

The obtained microscopy pictures were analysed using ImageJ software to extract data for $R$, $R_c$, $U_S$ and $U_F$. For measurements of the fibres extrusion speed, $U_F$, the build-in segmented-line command was used and the fibres were outlined manually. The fibre extrusion speed was calculated as the slope of the newly extruded fiber length per unit time, see Supplementary Figure S1. For measurement of the swimming speed, $U_S$, MTrackJ plugin was used which allows an easy track of the versus time, see **Video 9**. Due to the complexity of the system, all measurements were performed manually.

16